\documentclass[nofootinbib,showpacs,preprintnumbers,amsmath,amssymb,floatfix]
{revtex4-1}
\usepackage{graphicx}
\usepackage{wrapfig}
\begin{document}

%\begin{center}

\title{Factorization model for distributions of quarks in hadrons}

\vspace*{0.3 cm}

\author{B.I.~Ermolaev}
\affiliation{Ioffe Physico-Technical Institute, 194021
 St.Petersburg, Russia}
 \author{M.~Greco}
\affiliation{Department of Mathematics and Physics and INFN, University Roma Tre,
Rome, Italy}
\author{S.I.~Troyan}
\affiliation{St.Petersburg Institute of Nuclear Physics, 188300
Gatchina, Russia}

\begin{abstract}
We consider distributions of unpolarized (polarized)
quarks in unpolarized (polarized) hadrons. Our approach is based on QCD factorization.
We begin with study of Basic factorization for the parton-hadron
scattering amplitudes in the forward kinematics and suggest a model for
non-perturbative contributions to such amplitudes.
This model is based on the simple observation: after emitting an active quark by
the initial hadron,
the remaining set of quarks and gluons becomes unstable, so description of
this colored state can approximately be done in terms of resonances,
which leads to expressions of the Breit-Wigner type.
Then we reduce
these formulae to obtain explicit expressions for the quark-hadron
scattering amplitudes and quark distributions in
$K_T$- and Collinear factorizations.
\end{abstract}


\maketitle

\section{Introduction}

\begin{figure}[h]
  \includegraphics[width=.35\textwidth]{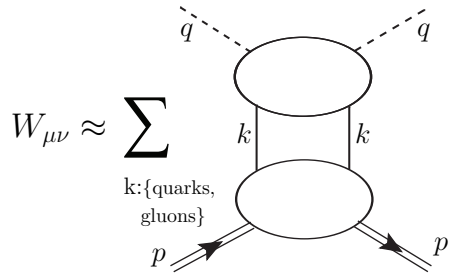}
  \caption{\label{pdfig1} Conventional illustration of QCD factorization. The $s$-cut
 of the graph is implied.}
\end{figure}

QCD factorization, i.e. separation of perturbative and
non-perturbative QCD contributions, proved to be an efficient
instrument for describing hadron reaction at high energies. Being
first applied to processes in the hard kinematics in the form
of Collinear factorization\cite{factcol}, it was soon extended to
cover the forward kinematic region, with DGLAP\cite{dglap} used to
account for perturbative contributions. Then, in order to be able
to use BFKL\cite{bfkl}, a new kind of factorization, $K_T$
-factorization was suggested in Ref.\cite{factkt}. These kinds of
factorization are usually illustrated by identical pictures. For
instance, factorization of the DIS hadronic tensor $W_{\mu\nu}$ is
conventionally depicted by the construction in Fig.~\ref{pdfig1}
both in Collinear and in $K_T$- factorizations, where the upper,
perturbative blob and the lower, non-perturbative blob are
connected by two-parton state.
%%%%%%%%%%%%%%%%%%%%%%%%%%%%%%%%%%%%%%%%%%%%
%%%%%%%%%%%%%%%%%%%%%%%%%%%%%%%%%%%%%%%%%%%
The upper blob in Fig.~\ref{pdfig1} is calculated with regular
perturbative means. On the contrary, the lower blob is
conventionally introduced from purely phenomenological
considerations. Collinear and $K_T$- factorizations operate with
different parametrizations for momentum $k$ of the connecting
partons and as a result, they are described by different formulae.
Collinear factorization assumes that

\begin{equation}\label{kcol}
%\vec{k} = \beta \vec{p},
k = \beta p,
\end{equation}

while $K_T$ -factorization allows for the transverse momentum in addition:

\begin{equation}\label{kkt}
%\vec{k} = \beta \vec{p} + \vec{k}_{\perp},
k = \beta p + k_{\perp},
\end{equation}
accounting therefore for one longitudinal and two transverse
components of $k$. However as a matter-of-fact, $k$ has four
components: two of them are longitudinal and the other two are
transverse. Accounting for the missing longitudinal component
$\alpha$ (for definition of $\alpha$ see Eq.~(\ref{sud})) drove us to suggesting a new, more
general factorization which we named in Ref.~\cite{egtfact} Basic
factorization. In contrast to $K_T$- and Collinear factorizations,
the analytic expressions in Basic factorization can be obtained from
the graphs of the type of the one in Fig.~\ref{pdfig1} with
applying the standard Feynman rules.

\begin{figure}[h]
\includegraphics[width=.15\textwidth]{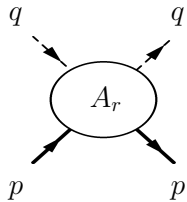}
\caption{\label{pdfig2} Amplitude for forward
Compton scattering off a hadron target.}
\end{figure}

It is worth reminding briefly our derivation of Basic factorization, for detail see Ref.~\cite{egtfact}.
Let us consider the Compton scattering amplitude off a hadron in the forward kinematics.
It is depicted in Fig.~\ref{pdfig2}.

The blob in Fig.~\ref{pdfig2} denotes ensemble
of perturbative and non-perturbative contributions. This blob can be expanded into an
infinite series of terms, each of them is represented by two blobs connected with
$n$ parton lines, $n = 2,3,..$. Considering only the simplest, two-parton state,
we arrive to the graph similar to the one in r.h.s of Fig.~\ref{pdfig1} but without the $s$-cut and
with the both blobs accommodating  perturbative and non-perturbative contributions at the same time.
The integration of the convolution in Fig.~\ref{pdfig1} over momentum $k$ now runs over the
whole phase space and it is expected to bring a finite result. However, the propagators
of the connecting partons become singular at $k^2 = 0$ (we neglect quark masses).
Besides, the upper blob may contain IR-sensitive perturbative contributions
$\sim \ln^n(2pk/k^2)$ (with $n= 1,2,..$). In addition, it yields the factor $2qk/k^2$, when
unpolarized gluon ladders are included into consideration. The only way to kill such IR singularity is to
assume that the lowest, non-perturbative blob should tend to zero fast enough when
$k^2 \to 0$ . Doing so and repeating a similar procedure to regulate the UV
singularity, we bring the convolution in  Fig.~\ref{pdfig1} to agreement with the
factorization concept: perturbative and non-perturbative contributions
are located indifferent blobs. This is a new form of QCD factorization which we name Basic
factorization.

We demonstrated in Ref.~\cite{egtfact}
that Basic factorization can be
reduced step-by-step first to $K_T$- and then to Collinear
factorizations. In Ref.~\cite{egtfact} we began with considering Basic factorization for
Compton scattering amplitudes in the forward kinematics, where
integration over momentum $k$ of the connecting partons in
Fig.~\ref{pdfig1} runs over the whole phase space. Confronting two
obvious facts that, on one hand, the integration over $k$ should
yield a finite results and that, on the other hand, the
perturbative part in Fig.~\ref{pdfig1}
(the upper, perturbative blob
and propagators of the connecting partons) is divergent in both the
infra-red (IR) and ultra-violet (UV) regions, allowed us to impose
integrability restrictions on the lowest blob, which are necessary for the
convolution in Fig.~\ref{pdfig1} to be finite.
%to kill the IR and UV divergences.
The obtained restrictions led us to theoretical constraints on the fits for the
parton distributions to the DIS structure functions in Collinear and $K_T$- factorizations.
In particular, we predicted the general
form of the fits in $K_T$ -factorization and excluded the factors $x^{-a}$ from the fits
in both $K_T$- and Collinear factorizations.

Another interesting object, where factorization is used, is distributions of partons
in hadrons. In the present paper we examine their properties in IR and UV regions and
suggest a simple resonance model for the non-perturbative contributions to the parton distributions.
Our argumentation in favor of this model is as follows: after emitting an active quark by
a hadron, the remains of the hadron, i.e. a set of quarks and gluons, acquires a
color and therefore it becomes unstable. So, this colored state can be described in
terms of resonances.
We begin with considering amplitudes of the quark-hadron (QHA) and gluon-hadron (GHA) scattering
in the forward kinematics. The Optical theorem relates such amplitudes to
the parton distributions.
Throughout the paper we use the standard Sudakov parametrization\cite{sud} for momentum $k$ of the connecting partons:

\begin{equation}\label{sud}
k= -\alpha q' + \beta p' + k_{\perp},
\end{equation}
where momenta $q'$ and $p'$ are massless, $p'^2 \approx q'^2 \approx 0$, and they are made of the hadron
momentum $p$ and the parton momentum $q$:

\begin{equation}\label{pq}
p' = p + x_2 q,~~q'= q + x_1 p,
\end{equation}
where $x_2 = -p^2/w \equiv - M^2/w,~ x_1 = -q^2/w$, with $w= 2pq \approx 2p'q'$. In these terms

\begin{equation}\label{pqk}
2pk = w(-\alpha - x_2 \beta),~~2qk = w(\beta - x_1 \alpha),~~ k^2 = -w \alpha \beta - k^2_{\perp}.
\end{equation}
In Sect.~II we introduce the quark-hadron scattering amplitudes in the forward
kinematics and examine their IR and UV behavior.
In Sect.~III we consider separately the unpolarized and spin-dependent
quark-hadron amplitudes in Basic factorization and suggest a model for non-pertubative contributions
to the amplitudes.
This model involves a spinor structure accompanied by invariant amplitudes
 $T^{(U)}$ and $T^{(S)}$.
In Sect.~III we  specify the spinor structure of the non-perturbative contributions
to the amplitudes and parton distributions. In Sect.~IV we
show how Basic factorization for the quark-hadron amplitudes and quark distributions
in hadrons can be reduced to $K_T$- and Collinear factorizations.
In Sect.~V we focus on a model for the invariant amplitudes $T^{(U)}$ and $T^{(S)}$.
The model is based on description of $T^{(U)}$ and $T^{(S)}$ in a quasi-resonant way
and through the Optical theorem it easily leads to non-perturbative contributions to the
parton distributions, with expressions of the Breit-Wigner kind both in Basic and in $K_T$- factorizations.
Finally, Sect.~VI is for concluding remarks.

  \section{Quark-hadron amplitudes}

\begin{figure}[h]
  \includegraphics[width=.25\textwidth]{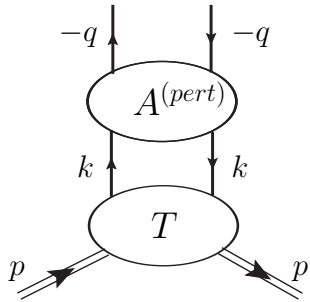}
  \caption{\label{pdfig3} Factorization of
the quark-hadron amplitude.}
\end{figure}

In the factorization approach, the quark-hadron amplitudes (QHA) $A_q$ are expressed through
convolutions of perturbative amplitudes $A^{(pert)}$ and non-perturbative
amplitudes $T$
%As , each of the factorized quark-hadron amplitudes (QHA) includes
%the lowest, non-perturbative blob $T$ and the perturbative blob
as shown in Fig.~\ref{pdfig3}.

%%%%%%%%%%%%%%%%%%%%%%%%%%%%%%%%%%%%%%%%%%%%
\begin{figure}[h]
  \includegraphics[width=.25\textwidth]{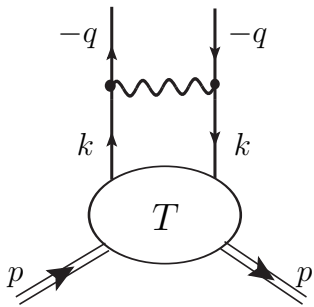}
  \caption{\label{pdfig4} Born approximation
for the factorized quark-hadron amplitude.}
\end{figure}

%%%%%%%%%%%%%%%%%%%%%%%%%%%%%%%%%%%%%%%%%%%
In the Born approximation $A^{(pert)}$ is depicted in Fig.~\ref{pdfig4} as a one-rung ladder.
Adding more ladder rungs to it together with inclusion of non-ladder graphs and resumming
all such graphs
converts the Born amplitude into $A^{(pert)}$. In the present paper we do not consider
mixing of quark and gluon ladder rungs, i.e. we consider the graphs where the vertical
quark lines go from the bottom to the top without breaking.

%%%%%%%%%%%%%%%%%%%%%%%%%%%%%%%%%%%%%%%%%%%%
%%%%%%%%%%%%%%%%%%%%%%%%%%%%%%%%%%%%%%%%%%%

We begin consideration of the quark-hadron amplitudes $A_q$ in Basic factorization, studying the simplest
case depicted in Fig.~\ref{pdfig4}, where the perturbative
contributions are accounted in the Born approximation and denote
such distributions $B_q$. In Basic factorization one can use the standard
Feynman rules to write down the analytic expression corresponding to the graphs in Figs.~3,4.
Doing so, we obtain that

\begin{equation}\label{bgen}
B_q = - \imath 4 \pi \alpha_s C_F \int \frac{d^4 k}{(2 \pi)^4}
\frac{\bar{u}(q)\gamma_{\mu}\hat{k}\hat{T}_q (k,p) \hat{k}\gamma_{\nu}u(q)}{k^2 k^2 (q+k)^2} d_{\mu\nu},
\end{equation}
where we have used the standard notations: $C_F = (N^2 -1)/(2N) =
4/3$ and $\alpha_s$ is the QCD coupling. In Eq.~(\ref{bgen})
$\hat{T}_q $ corresponds to the lowest blob in Fig.~\ref{pdfig3}.
It is altogether non-perturbative object. Throughout the paper we
will address it as the primary quark-hadron amplitude\footnote{In Ref.~\cite{brod} non-perturbative
contributions to parton distributions in the context of Collinear factorization were called intrinsic contributions.}. Choosing
the Feynman gauge, where $d_{\mu\nu} =  g_{\mu\nu}$, for the
virtual gluon and the Sudakov parametrization (\ref{sud}) for the
quark momentum $k$, we rewrite Eq.~(\ref{bgen}) as follows:

\begin{equation}\label{bsud}
B_q = - \imath \frac{\alpha_s C_F}{8 \pi^3} w \int d\alpha d \beta d^2 k_{\perp}
\frac{\bar{u}(q)\gamma_{\mu}\hat{k}\hat{T}_q (k,p) \hat{k}\gamma_{\mu}u(q)}{k^2 k^2 (q+k)^2}.
\end{equation}

Throughout the paper, for the sake of simplicity, we will treat the external quarks with momentum $q$
as on-shell ones, though our reasoning remains valid also when they are off-shell. Introducing the density
matrix

\begin{equation}\label{roq}
\hat{\rho}(p) (q)= \frac{1}{2} (\hat{q} + m_q)(1 - \gamma_5 \hat{S}_q)
,
\end{equation}
with $q,~m_q$ and $S_q$ being the quark momentum, mass and spin respectively,
we bring Eq.~(\ref{bsud}) to the following form:

\begin{equation}\label{bsud1}
B_q \approx  - \imath \frac{\alpha_s C_F}{8 \pi^3} w \int d\alpha d \beta d^2 k_{\perp}
\frac{Tr \left[\hat{\rho}(p) (q) \gamma_{\mu}\hat{k}\hat{T}_q (k,p) \hat{k}\gamma_{\mu}\right]}{k^2 k^2 (q+k)^2}.
\end{equation}
We stress that the replacement of Eq.~(\ref{bsud}) by Eq.~(\ref{bsud1}) is not necessary for us but
it allows us to carry out a more detailed consideration of $A_q^B$. In particular, we can consider separately
the spin-dependent, $B_q^{(spin)}$ and independent, $B_q^{(unpol)}$ quark-hadron amplitudes in a simple way:

\begin{equation}\label{bqunpol}
B_q^{(unpol)} = - \imath \frac{\alpha_s C_F}{8 \pi^3} w \int d\alpha d \beta d^2 k_{\perp}
\frac{2(qk) Tr\left[\hat{k}\hat{T}_q^{(unpol)}\right] - k^2Tr\left[\hat{q}\hat{T}_q^{(unpol)}\right] }{k^2 k^2 (q+k)^2},
\end{equation}

\begin{equation}\label{bqspin}
B_q^{(spin)} = \frac{\alpha_s C_F}{8 \pi^3}  m_q w \int d\alpha d \beta d^2 k_{\perp}
\frac{2(S_qk)~ Tr\left[\gamma_5\hat{k}\hat{T}_q^{(spin)}\right] - k^2Tr\left[\gamma_5\hat{S}_q\hat{T}_q^{(spin)}\right] }{k^2 k^2 (q+k)^2}.
\end{equation}
In Eqs.~(\ref{bqunpol},\ref{bqspin}) we have replaced the general primary amplitude $\hat{T}_q$ by
more specific amplitudes $\hat{T}_q^{(unpol)}, \hat{T}_q^{(spin)}$. In
Eq.~(\ref{bqunpol}) we have neglected a contribution $\sim m$ in $\hat{\rho} (q)$ compared to the
contribution $\sim \hat{q}$.
Integrations in Eqs.~(\ref{bqunpol},\ref{bqspin}) run over the whole phase space and it is supposed to yield finite results.
However, there can be singularities in the integrands and they should be regulated. Regulating them with
introducing various cut-offs would be unphysical, so
the only way out is to impose
appropriate constraints on the primary quark-hadron
amplitudes $\hat{T}_q^{(unpol)}, \hat{T}_q^{(spin)}$ so that to kill the singularities. When the perturbative
amplitude $A^{(pert)}$ is calculated in the Born approximation, the only possible
singularities in Eqs.~(\ref{bqunpol},\ref{bqspin}) are IR singularities at $k^2 = 0$
and UV singularities which  we relate to integrations over $\alpha$. However, when
$A^{(pert)}$ is beyond the Born approximation, there appears another kind of singularities
called in  Ref.~\cite{collinsrapid} rapidity divergences. Below we consider handling these
singularities in the framework of Basic factorization.

\subsection{Rapidity divergences of QHA}

\begin{figure}[h]
  \includegraphics[width=.5\textwidth]{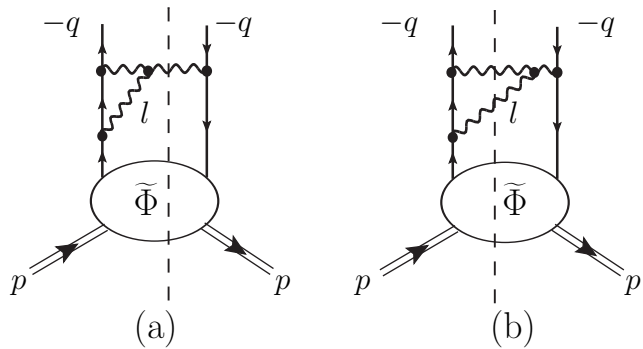}
  \caption{\label{pdfig5} Graphs contributing to rapidity divergences in unintegrated parton distributions.
The dashed lines denote cuts.}
\end{figure}

Rapidity divergences were investigated first in  Ref.~\cite{collinsrapid}
and then in Ref.~\cite{cheredrapid} in the context of $K_T$ -factorization. Detailed investigation of this problem
 can be found in Ref.~\cite{sterm}. In the lowest order of the Perturbative QCD,
the rapidity divergences come from the graphs in Fig.~\ref{pdfig5} (and symmetrical graphs as well), where the radiative corrections calculated in the
first-loop approximation are convoluted with the unintegrated parton distribution $\widetilde{\Phi}$.
Let us stress that $\widetilde{\Phi}$ accumulates both perturbative and non-perturbative corrections.
%%%%%%%%%%%%%%%%%%%%%%%%%%%%%%%%%%%%%%%%%%%%
%%%%%%%%%%%%%%%%%%%%%%%%%%%%%%%%%%%%%%%%%%%
%The vertical dashed lines in Fig.~\ref{pdfig5} correspond to the $s$-cuts.
When
such convolutions are considered in $K_T$ -factorization, each of the graphs in Fig.~\ref{pdfig5} acquires
logarithmic divergences arising from integration over momentum $l_+$ (with $l_+ = (l_0 + l_z)/\sqrt{2}$).
%providing $l^2 = 0$.
They are called rapidity divergences
and they can be got rid of as shown in Ref.~\cite{collinsrapid} (when the Feynman gauge is used for the gluon propagators) and then in
Ref.~\cite{cheredrapid} for the case of the light-cone  gauge.
In Refs.~\cite{collinsrapid,cheredrapid}
%is supposed that the lowest blob $\widetilde{\Phi}$ in $K_T$ -factorization
the rapidity divergences are cured with
redefining $\widetilde{\Phi}$.

Now let us study this situation in Basic factorization. To this end we consider a contribution of the graph in Fig.~6
to the quark-hadron amplitude in Basic factorization. We remind that there are no cuts in Fig.~6 and the blob $T$
accumulates non-perturbative contributions only.
%%%%%%%%%%%%%%%%%%%%%%%%%%%%%%%%%%%%%%%%%%%%
\begin{figure}[h]
  \includegraphics[width=.5\textwidth]{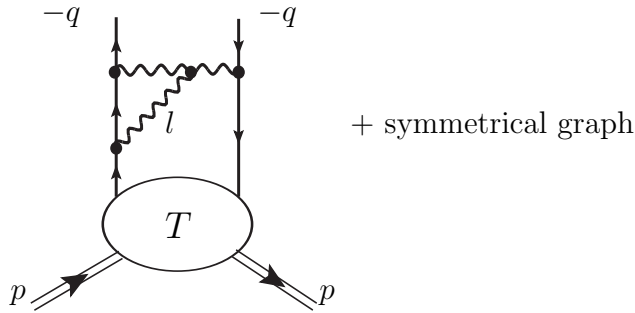}
  \caption{\label{pdfig6} Graph contributing to quark-hadron amplitude  .}
\end{figure}
%%%%%%%%%%%%%%%%%%%%%%%%%%%%%%%%%%%%%%%%%%%
One of remarkable features here is
that analytic expressions in Basic factorization can be obtained by applying standard Feynman
rules to the involved graphs. Second important point is that one is free to use any gauge
for perturbative QCD calculations\footnote{For gauge invariance of Basic factorization see Ref.~\cite{egtfact}. } in
Basic factorization whereas the blob $T$ in Fig.~6 is altogether non-perturbative
and therefore it is insensitive to the choice of the gauge.
Applying the Feynman rules to the graph in Fig.~\ref{pdfig6} and integrating over the loop momentum
$l$, we immediately conclude that
this integration yields a logarithmic UV-divergent contribution which, being complemented by a similar
contribution from
the symmetrical graph and self-energy graphs, in a
conventional way leads to renormalization of the gluon-quark couplings. After absorption of
such divergent contributions by the couplings, we obtain a renormalized amplitude which is free of
divergences. Then, applying the Optical theorem to the this construction, we arrive at
the parton distributions and they are also free of divergences.
Obviously, the same treatment can be applied to other UV divergences coming from perturbative
component $A^{(pert)}$ in higher loops:
all of them can be absorbed by renormalizations.
 Now we focus on the divergences resulting from integration of
the convolutions in Eqs.~(\ref{bqunpol},\ref{bqspin}), where the perturbative
amplitudes $A^{(pert)}$ are in the Born approximation.

\subsection{IR and UV stability of QHA}

First of all, let us note that the denominators in Eqs.~(\ref{bqunpol},\ref{bqspin}) can become singular in the infra-red (IR) region, where $k^2 \sim 0$.
In the case of purely perturbative QCD, IR singularities are conventionally regulated by introducing IR cut-offs.
In our case there is not any physical reason for that, so we are left with the only way to kill
these singularities: The primary quark-hadron amplitudes $\hat{T}_q$ should become small at small $k^2$:

\begin{equation}\label{tqir}
\hat{T}_q^{(unpol)}, \hat{T}_q^{(spin)} \sim \left(k^2\right)^{1+ \eta},
\end{equation}
when $k^2 \to 0$.
%This requirement coincides with the one in Eq.~(\ref{tir}).
%Substituting Eq.~(\ref{dqtir}) in Eqs.~(\ref{aqunpol},\ref{aqspin})
%leads us to
%
%\begin{equation}\label{aqunpol1}
%B_q^{(unpol)} =  \frac{\alpha_s C_F}{8 \pi^3} w \int d\alpha d \beta d^2 k_{\perp} (k^2)^{\eta}~
%\frac{2qk Tr\left[\hat{k}\hat{\textbf{T}}_q\right] - k^2Tr\left[\hat{q}\hat{\textbf{T}}_q\right] }{k^2 (q+k)^2},
%\end{equation}
%
%\begin{equation}\label{aqspin1}
%B_q^{(spin)} = \frac{\alpha_s C_F}{8 \pi^3} m_q w \int d\alpha d \beta d^2 k_{\perp}(k^2)^{\eta}~
%\frac{2S_qk Tr\left[\gamma_5\hat{k}\hat{\textbf{T}}_q\right] - k^2Tr\left[\gamma_5\hat{S}_q\hat{\textbf{T}}_q\right] }{k^2 (q+k)^2}.
%\end{equation}
%
%\subsection{UV stability of QHA}
Now let us consider the ultra-violet (UV) stability of the convolutions in Eqs.~(\ref{bqunpol},\ref{bqspin}).
The integration over $\alpha$ in Eqs.~(\ref{bqunpol},\ref{bqspin}) runs from $- \infty$ to $\infty$,
so, at large $|\alpha|$ the integrands should decrease fast enough to guarantee UV stability.
First of all we focus on the integration over $\alpha$ in Eq.~(\ref{bqunpol}).
Taking into consideration that each factor in the denominator
 of Eq.~(\ref{bqunpol}) is $\sim \alpha$ makes that the denominator to be $\sim \alpha^3$.
The term $2qk$ in the numerator depends on $\alpha$ because
$2qk = w(\beta - x_1 \alpha)$ and the factors $k^2$ and $\hat{k}$ are $\sim \alpha$, which makes

\begin{equation}\label{bunpola}
\frac{2qk~ \hat{k}}{k^2 k^2 (q+k)^2} \sim \frac{\alpha^2}{\alpha^3}.
\end{equation}
This divergence must be regulated by an appropriate decrease of $\hat{T}_q^{(unpol)}$ at large $|\alpha|$.
The IR stability condition in Eq.~(\ref{tqir}) states that $\hat{T}_q^{(unpol)} \sim \left(k^2\right)^{1 + \eta}$
at small $k^2$ but it can either disappear or be kept at large $|\alpha|$. Therefore we have two options:\\
\textbf{(A)} The factor $\left(k^2\right)^{1 + \eta}$ survives at large $|\alpha|$.\\
\textbf{(B)} The factor $\left(k^2\right)^{1 + \eta}$ disappears at large $|\alpha|$.\\
In the case \textbf{(A)}, where IR and UV behaviors of $\hat{T}_q^{(unpol)}$ are related,
$\hat{T}_q^{(unpol)}$ should behave at large $|\alpha|$ as follows:

\begin{equation}\label{tqunpol}
\hat{T}_q^{(unpol)} \sim \alpha^{\eta -\chi} = \left(\alpha^{1 + \eta}\right) \left[\alpha^{-1 - \chi}\right],
\end{equation}
with $\chi > \eta > 0$. \\
IR and UV behaviors of $\hat{T}_q^{(unpol)}$ are disconnected in the case \textbf{(B)}. It converts Eq.~(\ref{tqunpol}) into

\begin{equation}\label{tqunpol1}
\hat{T}_q^{(unpol)} \sim \alpha^{-\chi}.
\end{equation}

The first factor in Eq.~(\ref{tqunpol}) corresponds to the term
$\left(k^2\right)^{1 + \eta}$, while a contribution generating  the asymptotic factor in the squared brackets has to
be specified. We will do it in Sect.~V.
Now let us consider the spin-dependent amplitudes. In order to guarantee their IR stability,
the primary spin-dependent amplitude $\hat{T}_q^{(spin)}$ should also be $\sim (k^2)^{1 + \eta}$
at small $k^2$ but the
situation with its UV stability is more involved than in the unpolarized case. Indeed,
the quark spin $S_q$ can be either in the plane formed by $p$ and $q$, i.e. $S_q = S_q^{||}$, or
in the transverse space, where $S_q = S_q^{\perp}$. Depending on it, there are the longitudinal spin-dependent
amplitude, $B_q^{||}$ and the transverse one, $B_q^{\perp}$. Now let us consider
the term $2m_q S_qk$ in Eq.~(\ref{bqspin}) for different orientations of the quark spin:
When the spin is longitudinal,

\begin{equation}\label{spinl}
2m_qS_q k  = 2m_qS_q^{||}k = w (\beta - x_1 \alpha)
\end{equation}

and $\hat{k}$  in the trace $Tr[\hat{k}\hat{T}_q]$ is also $\sim \alpha$.
In contrast when the spin is transverse,
\begin{equation}\label{spint}
2m_qS_q k = -2m_q (\vec{S}_q^{\perp} \vec{k}_{\perp})
\end{equation}
%So, $S_q^{||}k \sim \alpha$ while
and therefore $S_q^{\perp}k$ does not depend on $\alpha$. Then,
this $\vec{k}_{\perp}$ should be accompanied by another $\vec{k}_{\perp}$ from
the trace in order to get a non-zero result at integration over the azimuthal angle,
i.e. The first term in the numerator of Eq.~(\ref{bqspin})
%i.e. $k$ in the first trace of Eq.~(\ref{bqspin}) is also transverse  and
does not depend on $\alpha$, while the second term is $\sim \alpha$.
It means that, with $\hat{T}_q^{||}$ dropped,
the explicit $\alpha$-dependence of $A_q^{||}$  at large $|\alpha|$ coincides with
the one in Eq.~(\ref{bunpola}):

\begin{equation}\label{uvspinl}
\frac{S_q^{||}k~\hat{k}}{k^2 k^2 (q+k)^2}\sim \frac{\alpha^{2}}{\alpha^3}
\end{equation}
and
\begin{equation}\label{uvspint}
\frac{S_q^{\perp}k~\hat{k}}{k^2 k^2 (q+k)^2} \sim \frac{\alpha}{\alpha^3}.
\end{equation}

It follows from Eq.~(\ref{uvspinl}) states that the $\alpha$ -dependence of
the amplitude $\hat{T}_q^{||}$
at large $|\alpha|$ is identical to the one of $\hat{T}_q^{(unpol)}$:

\begin{equation}\label{tqspinl}
\hat{T}_q^{||} \sim \hat{T}_q^{(unpol)}  \sim \left(\alpha^{1 + \eta}\right) \left[\alpha^{-1 - \chi}\right]
\end{equation}
in the case \textbf{(A)} and
\begin{equation}\label{tqspinl1}
\hat{T}_q^{||} \sim \hat{T}_q^{(unpol)}  \sim \left(\alpha^{-\chi}\right)
\end{equation}
in the case \textbf{(B)}.
$\hat{T}_q^{\perp}$ can decrease slower:
\begin{equation}\label{tqspint}
\hat{T}_q^{\perp} \sim \left(\alpha^{1 + \eta}\right) \left[\alpha^{-\chi}\right]
\end{equation}
in the case \textbf{(A)} and
\begin{equation}\label{tqspint1}
\hat{T}_q^{\perp} \sim \alpha^{1-\chi}
\end{equation}
in the case \textbf{(B)}.
Eqs.~(\ref{tqir},\ref{tqspinl} - \ref{tqspint1}) guarantee integrability of the convolutions for the quark-hadron
amplitudes in Basic factorization. 
These integrability requirements can be used as general theoretical constraints on non-perturbative
contributions to the amplitudes in Basic factorization (see Ref.~\cite{egtfact} for detail) and we will use 
them  
in the present paper.    
%When (we consider it in Sect.~IV) Basic factorization is reduced to $K_T$ and Collinear factorization,
%the theoretical constraints are also modified  and can be used as
%constraints on the fits for unintegrated and integrated parton distributions.
Each of Eqs.~(\ref{tqspinl},\ref{tqspint}) consists of two factors. The first factor in these equations is universally
generated by the term $\left(k^2\right)^{1+ \eta}$ while
contributions generating the factors in squared brackets will be specified in Sect.~V.
%As soon as the parameters $\eta, \chi$ are arbitrary, it seems that there is no difference between the decreases of
%$\hat{T}_q^{||}$ and $\hat{T}_q^{\perp}$ in
%Eqs.~(\ref{tqspinl},\ref{tqspint}). However, such a difference will appear, when we choose $\eta, \chi$
%to be small.

\section{Modeling the spinor structure of $\hat{T}_q$}

Our next step is to simplify the traces in Eqs.~(\ref{bqunpol},\ref{bqspin}).
In order to do it, we have to specify the spinor structure of the
primary QHA $\hat{T}_q$.
By definition, $\hat{T}_q$ is altogether non-perturbative, so
specifying its spinor structure can only be done on basis of phenomenological considerations.
However, any model expression for $\hat{T}_q$ should respect the integrability conditions in
Eqs.~(\ref{tqir}, \ref{tqunpol},\ref{tqspinl},\ref{tqspint}).
There is the well-known expression for the density matrix of an elementary fermion:
\begin{equation}\label{rop}
\hat{\rho}(p) = \frac{1}{2} (\hat{p} + M)(1 - \gamma_5 \hat{S}) \approx \frac{\hat{p}}{2} -
\frac{1}{2} (\hat{p} + M) \gamma_5 \hat{S},
\end{equation}
where $M$ and $S$ are the fermion mass and spin. This expression drives us to
approximate $\hat{T}_q$ as follows:

\begin{equation}\label{tus}
\hat{T}_q =  \hat{p}~ T_q^{(U)} (k^2, 2pk) - (\hat{p} + M) \gamma_5 \hat{S}~ T_q^{(S)} (k^2, 2pk),
\end{equation}
where $p,~S$ are the hadron momentum and spin respectively and $T_q^{(U)},~T_q^{(S)}$ are scalar functions.
Throughout the paper we will address them as invariant quark-hadron amplitudes.
Substituting $T_q $ of Eq.~(\ref{tqir}) in Eqs.~(\ref{bqunpol},\ref{bqspin}) and calculating the traces, we arrive
at the following expressions:

\begin{eqnarray}\label{bqunpolt}
B_q^{(unpol)} &=& -\imath \frac{1}{8\pi^3} \int d\alpha d \beta d k^2_{\perp}
\left[- g^2 C_F \frac{w  }{ \left[(q+k)^2 + \imath \epsilon\right]}\right]
\left(\frac{ k^2_{\perp}}{k^2 k^2}\right)
\left(T_q^{(U)}(k^2,2pk)\right)
\\ \nonumber
&=& -\imath \frac{1}{8\pi^3} \int d\alpha d \beta d k^2_{\perp}
\widetilde{B}_q^{(unpol)} (q,k)
\left(\frac{ k^2_{\perp}}{k^2 k^2}\right)
\left(T_q^{(U)}(k^2,2pk)\right),
\end{eqnarray}
were we have denoted $\widetilde{B}_q^{(unpol)}$ the perturbative  amplitude in the Born approximation
for the forward annihilation of unpolarized quark-quark pair.
We have neglected contributions $\sim x_{1,2}$ in the numerator of Eq.~(\ref{bqunpolt}) and will
do it in expressions for the spin-dependent amplitudes.
These terms, if necessary, can easily be accounted for with more accurate implementation of Eq.~(\ref{sud}) to Eqs.~(\ref{bqunpolt}).
Let us consider the structure of the integrand in Eq.~(\ref{bqunpolt}) in more detail.
The amplitude in the
last brackets is entirely non-perturbative. It is suppose to mimic a transition from hadrons to quarks.
The fraction in the middle corresponds to the convoluting the perturbative and non-perturbative amplitudes.
%Eq.~(\ref{dqspin1}) has the same structure.
The fraction in the first brackets
corresponds to the perturbative amplitude for the forward scattering of quarks in
the Born approximation. We explicitly wrote the factor
$\imath \epsilon$ there to remind that this amplitude has the $s$-channel imaginary part. Doing similarly,
we obtain an expression for the spin-dependent amplitudes:

\begin{eqnarray}\label{aqspin2}
B_q^{(spin)} &=& \imath \frac{g^2 C_F}{16\pi^4} 2 m_q M w \int d\alpha d \beta d^2 k_{\perp}
\frac{2(kS_q)( kS) - k^2 (S_qS) }{k^2 k^2 \left[(q+k)^2 + \imath \epsilon\right]}~T_q^{(S)}
%\\ \nonumber
\end{eqnarray}

Let us consider Eq.~(\ref{aqspin2})  for different orientation of the hadron spin: \\
(\textbf{i}) The hadron spin $S$ is in the plane formed by momenta $p$ and $q$, so for this
case we use the notation $S = S^{||}$.\\
(\textbf{ii}) The hadron spin is transverse to this plane. We denote this case as $S = S^{\perp}$.

Amplitude $A_q^{\|}$ for the first case is given by the expression very close to the unpolarized amplitude:

\begin{eqnarray}\label{bl}
B^{(\|)}_q &=& - \imath \frac{1}{16 \pi^3 } \int d\alpha d \beta d k^2_{\perp}
\left[-g^2 C_F \frac{2 m M (S_q^{\|} S^{\|})}{(q+k)^2 + \imath \epsilon}\right]
\left(\frac{ k^2_{\perp}}{k^2 k^2}\right) T_q^{(\|)}(k^2,2pk)
\\ \nonumber
&=& - \imath \frac{1}{8\pi^4} \int d\alpha d \beta d^2 k_{\perp}
\widetilde{B}_q^{(\|)} (q,k)
\left(\frac{ k^2_{\perp}}{k^2 k^2}\right) T_q^{(\|)}(k^2,2pk),
\end{eqnarray}
whereas the transverse amplitude is given by a different expression:

\begin{eqnarray}\label{bt}
B^{(\perp)}_q &=&  - \imath \frac{1}{16 \pi^3 } \int d\alpha d \beta d k^2_{\perp}
\left[-g^2 C_F \frac{2 m M (S_q^{\perp} S^{\perp})}{(q+k)^2 + \imath \epsilon}\right]
\left(\frac{w \alpha \beta}{k^2 k^2}\right) T_q^{(\perp)}(k^2,2pk)
\\ \nonumber
&=& - \imath \frac{1}{8 \pi^3 } \int d\alpha d \beta d k^2_{\perp}
\widetilde{B}_q^{(\perp)} (q,k)
\left(\frac{w \alpha \beta}{k^2 k^2}\right) T_q^{(\perp)}(k^2,2pk) ,
\end{eqnarray}
with $\widetilde{B}_q^{(\|)} ,~\widetilde{B}_q^{(\perp)}$ being the perturbative spin-dependent Born amplitudes.
Accounting for perturbative QCD radiative corrections converts the Born amplitudes
$\widetilde{B}_q^{(unpol)}, \widetilde{B}_q^{(\|)}, \widetilde{B}_q^{(\perp)}$
in  Eqs.~(\ref{bqunpolt},\ref{bl},\ref{bt}) into perturbative dimensionless amplitudes
$\widetilde{A}_q^{(unpol)}, \widetilde{A}_q^{(\|)}, \widetilde{A}_q^{(\perp)}$, remaining the other factors unchanged:

\begin{eqnarray}\label{atot}
A_q^{(unpol)} (p,q) &=& -\imath \frac{1}{8\pi^3} \int
%\frac{d \beta}{\beta}
d \beta \frac{d k^2_{\perp}}{k^2} d\alpha
\widetilde{A}_q^{(unpol)} (q,k)
\left(\frac{ k^2_{\perp}}{k^2}\right)T_q^{(U)}(k^2,2pk),
\\ \nonumber
A^{(\|)}_q (p,q) &=&
- \imath \frac{1}{8\pi^3} \int
%\frac{d \beta}{\beta}
d \beta \frac{d k^2_{\perp}}{k^2} d\alpha
\widetilde{A}_q^{(\|)} (q,k)
\left(\frac{ k^2_{\perp}}{k^2}\right) T_q^{(\|)}(k^2,2pk),
\\ \nonumber
A^{(\perp)}_q (p,q) &=&  - \imath \frac{1}{8 \pi^3 } \int
%\frac{d \beta}{\beta}
d \beta \frac{d k^2_{\perp}}{k^2} d\alpha
\widetilde{A}_q^{(\perp)} (q,k)
\left(\frac{w \alpha \beta}{k^2}\right) T_q^{(\perp)}(k^2,2pk) .
\end{eqnarray}

Taking the $s$-imaginary part of Eq.~(\ref{bqunpolt}),
we arrive at the totally unintegrated, or fully unintegrated as was used in Ref.~\cite{collinsfully},
distribution of unpolarized quarks in the hadron $D_q^{unpol~B}$ in the Born approximation:

\begin{eqnarray}\label{dqunpolb}
D_q^{(unpol~B)} &=& \frac{1}{8\pi^2} \int d \beta \frac{d k^2_{\perp}}{k^2} d\alpha~
\left[g^2 C_F \delta \left(\beta - x - z\right) \right]\frac{k^2_{\perp}}{k^2} \Psi _q^{(1)}(k^2,2pk)
\\ \nonumber
&=& \frac{1}{8\pi^2} \int \frac{d \beta}{\beta} \frac{d k^2_{\perp}}{k^2} d\alpha~
\left[g^2 C_F \delta \left(1 - x/\beta - z/\beta\right) \right]\frac{k^2_{\perp}}{k^2} \Psi _q^{(1)}(k^2,2pk)
\end{eqnarray}
where $x = -q^2/w,~z = k^2_{\perp}/w$ and $\Psi_q^{(1)}$ is
the primary quark distribution of unpolarized quarks in the hadron,
$\Psi_q^{(1)}= (1/\pi) \Im T_q^{(U)}$. This object is altogether non-perturbative.
Applying the Optical theorem to Eq.~(\ref{atot}), we arrive at the
parton distributions beyond the Born approximation:

\begin{eqnarray}\label{dtot}
D_q^{(unpol)}(x, q^2) &=& \frac{1}{8\pi^2} \int
%\frac{d \beta}{\beta}
\frac{d \beta}{\beta} \frac{d k^2_{\perp}}{k^2} d\alpha
\widetilde{D}_q^{(unpol)} (x/\beta,q^2/k^2)
d \beta \left(\frac{ k^2_{\perp}}{k^2}\right) \Psi_q^{(1)}(k^2, w\alpha),
\\ \nonumber
D^{(\|)}_q (x, q^2) &=&
\frac{1}{8\pi^2} \int
\frac{d \beta}{\beta}
 \frac{d k^2_{\perp}}{k^2} d\alpha
\widetilde{D}_q^{(\|)} (x/\beta,q^2/k^2)
\left(\frac{ k^2_{\perp}}{k^2}\right) \Psi_q^{(\|)}(k^2, w\alpha),
\\ \nonumber
D^{(\perp)}_q (x, q^2) &=& \frac{1}{8 \pi^2 } \int
\frac{d \beta}{\beta}
\frac{d k^2_{\perp}}{k^2} d\alpha
\widetilde{D}_q^{(\perp)} (x/\beta,q^2/k^2)
\left(\frac{w \alpha \beta}{k^2}\right) \Psi_q^{(\perp)}(k^2, w\alpha) .
\end{eqnarray}

\section{Reduction of Basic factorization to conventional factorizations}

Conventional forms of factorization are Collinear and $k_T$- factorizations. In Ref.~\cite{egtfact}
we described reduction of Basic factorization to $K_T$- and Collinear factorizations for the Compton
scattering amplitudes and DIS structure functions
without specifying the non-perturbative amplitudes $T_q$. In this Sect. we show that
these results perfectly agree with our
assumption in Eq.~(\ref{tus}) concerning structure of $T_q$.  We
demonstrate that the parton
distributions in both conventional factorizations
can be obtained with step-by-step reductions of the expressions for
$D_q^{(unpol)}, D^{(\|)}_q, D^{(\perp)}_q$  in Basic factorization. This reductions are the same for
both the
parton-hadron amplitudes and parton distributions, they are insensitive to spin
and stands when the quarks are replaced by gluons. Because of that
we consider such reductions for a generic parton-hadron distribution $D$ in Basic factorization
and skip unessential factors:

\begin{equation}\label{dgen}
D (x, q^2_{\perp}) = \int
%\frac{d \beta}{\beta}
d \beta \frac{d k^2_{\perp}}{k^2} d\alpha
D^{(pert)} (x/\beta, q^2_{\perp}/k^2) \left(\frac{ k^2_{\perp}}{k^2}\right) \Psi (w\alpha, k^2),
\end{equation}
where $D^{(pert)}$ stands for a perturbative contribution and $\Psi$ is the
altogether non-perturbative (we address it as primary) parton-hadron distribution.
Actually, $\Psi (w\alpha, k^2)$ is the starting point for the perturbative evolution.
Integration in Eq.~(\ref{dgen}) runs over the whole phase space.
Let us note that in the literature very often are considered purely transverse $q$:
%usually are considered negative $q^2$, i.e.
$q^2 \approx q^2_{\perp}$. Because of this reason we will
use the notation $q^2_{\perp}$ instead of $q^2$ in what follows,
though Eqs.~(\ref{dtot},\ref{dgen}) are also valid when $q^2 \neq q^2_{\perp}$.

\subsection{Reduction to $k_T$ -factorization}

In order to reduce Eq.~(\ref{dgen}) to $k_T$ -factorization, we have to
perform integration with respect to $\alpha$. However, this integration should not involve
$D^{(pert)}$, which, strictly speaking, is impossible because $D^{(pert)}$ depends on $k^2$ and
thereby it depends on $\alpha$: $k^2 = - w \alpha \beta - k^2_{\perp}$. The only way out is to assume that the main contributions
to Eq.~(\ref{dgen}) come from the region where

\begin{equation}\label{abk}
\alpha \ll \alpha_{\max} = k^2_{\perp}/(w \beta),
\end{equation}
i.e. $k^2 \approx - k^2_{\perp}$. Let us notice that approximating ladder partons
virtualities $k^2$ by their transverse momenta is well-known. It is used in all available
evolution equations, including DGLAP and BFKL, and now it allows us to convert Eq.~(\ref{dgen}) into
an expression for the unintegrated (transverse momentum dependent\cite{collinstmd}) parton
distributions $D_{KT}$ in  $k_T$ -factorization:

\begin{equation}\label{dkt}
D_{KT} \left(x, q^2_{\perp}\right)\approx \int_x^1
\frac{d \beta}{\beta} \int_0^r
\frac{d k^2_{\perp}}{k^2_{\perp}}
D^{(pert)}_{KT} (x/\beta, q^2_{\perp}/k^2_{\perp}) \Phi (w\beta, k^2_{\perp}),
\end{equation}
where $r = w \beta - q^2_{\perp}$.
Obviously, $r \approx w \approx q^2_{\perp}$ for large $x$  while at very small $x$
one can use $r \approx w \beta$.
$\Phi$ denotes the primary (i.e. non-perturbative) $k_T$ -parton distribution. It is related to $\Psi$ as follows:

\begin{equation}\label{phidkt}
\Phi (w\beta, k_{\perp}) = \int_0^{k^2_{\perp}/w \beta} d \alpha \Psi (w \alpha, k^2_{\perp}).
\end{equation}

%As we do not study perturbative parts of the parton distributions in the present paper,
%We have used in Eqs.~(\ref{dgen},\ref{dkt}) the generic notation $D^{(pert)}$ for the perturbative
%contributions in $K_T$ -factorization and will keep this notation in the context of
%Collinear factorization.

\subsection{Reduction to Collinear factorization}

In Ref.~\cite{egtfact} we discussed how to reduce $K_T$ -factorization to Collinear one, using DIS
structure functions as an example. The same argumentation can be applied to the parton distributions.
We briefly repeat it below. In order to reduce $k_T$ -factorization to Collinear factorization,
we should perform integration of Eq.~(\ref{dkt}) with respect to
$k_{\perp}$ without integrating $D^{(pert)} $. Of course it cannot be done straightforwardly,
because  $D^{(pert)} $ explicitly depends on $k_{\perp}$. However, we can do it
approximately, assuming a sharp peaked dependence of $\Psi (w \alpha,
k^2_{\perp})$ on $k^2_{\perp}$ with maximum at $k^2_{\perp} =
\mu^2$ as shown in Fig.~\ref{pdfig7}. The close this dependence is to $\delta (k^2_{\perp} - \mu^2)$,
the higher is accuracy of the reduction. As discussed in Ref.~\cite{egtfact}, the number of such maximums
can be unlimited. We remind that $\Phi$ is non-perturbative, so typical values of $\mu$ must be of
non-perturbative range, $\mu \sim \Lambda_{QCD}$.
After the integration of $\Phi$ we arrive at Collinear factorization convolution:

%%%%%%%%%%%%%%%%%
\begin{figure}[h]
\includegraphics[width=.3\textwidth]{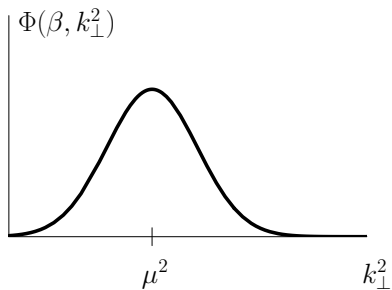}
%\begin{center}
%\begin{picture}(240,100)
%\put(0,0){\epsfxsize=50mm \epsfbox{pdfig7.eps} }
%\end{picture}
%\end{center}
\caption{\label {pdfig7} The peaked form of $\Phi(\beta,k_{\perp}^2)$ with one maximum}.
\end{figure}
%%%%%%%%%%%%%%%%%%

\begin{equation}\label{dcol}
D^{(col)} (x, q^2_{\perp}/\mu^2) \approx \int_x^1
\frac{d \beta}{\beta}
%D^{(pert)} (q^2, w\beta) \phi (x/w\beta, \mu^2)
%= \int d \beta
%\frac{d \beta}{\beta}
D^{(pert)}_{col} (x/\beta, q^2_{\perp}/\mu^2) \phi (\beta, \mu^2),
\end{equation}
with $\mu$ being the intrinsic factorization scale and $\phi$ being the primary (non-perturbative) integrated parton distribution:

\begin{equation}\label{phicolint}
\phi (\beta, \mu^2) = \int_{\Omega} \frac{d k^2_{\perp}}{k^2_{\perp}}
\Phi (w\beta, k^2_{\perp}),
\end{equation}
where the integration region $\Omega$ is located around the maximum $k^2_{\perp} = \mu^2$.  At the first sight, the form of Collinear
factorization presented in Eq.~(\ref{dcol}) contradicts to the conventional form.
Indeed, the scale $\mu$ in Eq.~(\ref{dcol}) corresponds to the maximum in Fig.~7 and therefore its value
is fixed. On the contrary, the conventional form of Collinear factorization operates with integrated
parton distributions  $\varphi(\beta,\widetilde{\mu}^2)$, where the scale $\widetilde{\mu}$ can have any arbitrary value. Then,
we expect the value of $\mu$ to be of non-perturbative range whereas usually
$\widetilde{\mu} \sim $ few GeV, i.e. typically $\widetilde{\mu} \gg \mu$.
However, this contradiction can
easily be solved as was shown in Ref.~\cite{egtfact}. The point is that transition from $ \phi (\beta, \mu^2)$ to
$\varphi(\beta,\widetilde{\mu}^2)$
%the intrinsic scale $\mu$ to an arbitrary scale $\widetilde{\mu}$
can be done, applying perturbative evolution in the $\mu^2$ -space to $ \phi (\beta, \mu^2)$ and keeping $\beta$ fixed. It can be written symbolically as

\begin{equation}\label{phimutilde}
\varphi(\beta,\widetilde{\mu}^2) = E (\widetilde{\mu}^2, \mu^2 )\otimes \phi (\beta, \mu^2),
\end{equation}
where $E$ is the evolution operator in the $\mu^2$ -space.
%The evolution operator $E$ is considered in detail in Appendix E.
Specific expressions
for $E$ are different in different perturbative approaches (see Appendix E for detail). Eq.~(\ref{phimutilde})
%Applying this evolution to the non-perturbative distribution $\phi (\beta, \mu^2)$
%allows us to change the intrinsic scale $\mu^2$ for
% (providing $\widetilde{\mu}> \mu$) and
makes possible to arrive at the conventional unintegrated distribution
$\varphi(\beta,\widetilde{\mu}^2)$ fixed at an arbitrary scale $\widetilde{\mu}^2$.
In contrast to $ \phi (\beta, \mu^2)$,
the distribution $\varphi(\beta,\widetilde{\mu}^2)$ accumulates both perturbative and non-perturbative contributions.
It is easy to show that our reasoning remains true in the case when $\Phi$ has several maximums or an infinite series of them.
This point was discussed in detail in Ref.~\cite{egtfact}, so
we will not do it in the present paper. Instead, we focus on modeling invariant amplitudes $T^{(U,S)}$ introduced in Eq.~(\ref{tus}).

\section{Modeling the invariant quark-hadron amplitudes and primary quark distributions }

In this Sect. we suggest a model which mimics non-perturbative QCD contributions in the
primary hadron-quark invariant amplitudes $T^{(U,S)}$ and in the primary quark distributions
in all available forms of factorization.
Once again we begin with consideration of the invariant amplitudes $T^{(U,S)}$ and then
proceed to the quark distributions.

\subsection{Resonance model for the primary quark-hadron invariant amplitudes}

Amplitudes $T^{(U,S)}$ can be introduced
in a model-dependent way only because QCD has not been solved in the non-perturbative region.
All such models should satisfy several restrictions: \\
\textbf{(i)}: The IR stability conditions in Eq.~(\ref{tqir})
and the UV stability conditions in Eqs.~(\ref{tqspinl} - \ref{tqspint1}) 
%in Eqs.~(\ref{tqunpol},\ref{tqunpol1},\ref{tqspinl},\ref{tqspinl1},\ref{tqspint},\ref{tqspint1})
should be respected because they guarantee integrability of the factorization convolutions. 
We remind that the UV stability conditions derived in Sect.~II depend on
UV-behavior of the factors regulating IR divergences . Namely,
Eqs.~(\ref{tqunpol},\ref{tqspinl},\ref{tqspint}) correspond to the case \textbf{(A)} while
Eqs.~(\ref{tqunpol1},\ref{tqspinl1},\ref{tqspint1}) correspond to the case \textbf{(B)}. In the present paper we
focus on the most UV-divergent case \textbf{(A)}, although our conclusions hold true for the case (B) as well.
 \\
\textbf{(ii)}: the invariant amplitudes should respect the Optical Theorem, so they should have $s$-channel
imaginary parts.\\
(\textbf{iii}): These amplitudes should guarantee the step-by-step reductions from Basic factorization to $k_T$ -factorization and then to
Collinear factorization described in Sect.~IV.\\
The expressions for the unpolarized and spin-dependent amplitudes with the longitudinal spin in Eq.~(\ref{atot})
are much alike while the expression for the transverse spin amplitude differs from them. Despite
this difference, our model equally stands for all spin-dependent amplitudes and quark
distributions regardless of the spin orientation.
%In what follows we consider the unpolarized quark-hadron amplitudes and
%the spin-dependent amplitudes with longitudinal spin. The case of transverse spin can be considered in a similar way.
To describe the invariant primary quark-hadron amplitudes, we suggest a model  of the resonance type for $T_q^{(U,S)}$:

\begin{eqnarray}\label{tusmod}
T_q^{(U)} (pk, k^2) &=&  \frac{R_U \left(k^2\right)}{\left((k- p)^2  - M^2_1 + \imath \Gamma_1 \right)
\left((k- p)^2  - M^2_2 + \imath \Gamma_2 \right)}
\\ \nonumber
T_q^{(S)} (pk, k^2) &=& \frac{
R_S\left(k^2\right)}{\left((k- p)^2  - M^2_3 + \imath  \Gamma_3\right)
\left((k- p)^2  - M^2_4 + \imath  \Gamma_4\right)}
\end{eqnarray}
where
$R_{U,S}\left(k^2\right)$ are supposed to behave as
$R_{U,S}\left(k^2\right) \sim \left(k^2\right)^{1+ \eta}$  at small $k^2$. We need at least two resonances to 
satisfy the UV stability requirement of Eq.~(\ref{tqunpol}). Indeed, Eq.~(\ref{tusmod}) leads to 
$\chi = 1$ while a model with one resonance corresponds to $\chi = 0$.
 Formally, Eq.~(\ref{tusmod}) contains
independent parameters $M_{1,2,3,4}$ and
$\Gamma_{1,2,3,4}$ but we do not see a physical reason forbidding to identify $T_q^{(U)}$ and $T_q^{(S)}$, 
which would left us with the parameters $M_{1,2}$ and
$\Gamma_{1,2}$ only.
%The factor $w^2$ is introduced from dimensional considerations.
In terms of the Sudakov variables $T_q^{(U,S)}$ are:

\begin{eqnarray}\label{tussud}
T_q^{(U)}(w \alpha, k^2) &=& \frac{R_U \left(k^2\right)}{\left(w \alpha + k^2 - \mu^2_1 + \imath  \Gamma_1\right)
\left(w \alpha + k^2 - \mu^2_2 + \imath  \Gamma_2\right)}
\\ \nonumber
T_q^{(S)}(w \alpha, k^2) &=& \frac{
R_S\left(k^2\right)}{\left(w \alpha + k^2 - \mu^2_3 + \imath  \Gamma_3\right)
\left(w \alpha + k^2 - \mu^2_4 + \imath  \Gamma_4\right)},
\end{eqnarray}
where $k^2 = - w \alpha\beta -k^2_{\perp}$ and

\begin{equation}\label{mu}
\mu^2_j = M^2_j - p^2.
\end{equation}
We suggest that values of $\mu^2_j$
and $\Gamma_j$
should be within the non-perturbative scale domain, with $M^2_j > \Gamma_j$.
It is convenient to write $T_{U,S}$ as the sum of two resonances:

\begin{eqnarray}\label{tussudalfa}
T_q^{(U)}(w \alpha, k^2)
&=& \frac{R_U \left(k^2\right)}{(\mu^2_1 - \mu^2_2) - \imath (\Gamma_1 - \Gamma_2)}
\left[\frac{1}{\left(w \alpha + k^2 - \mu^2_1 + \imath  \Gamma_1\right)}
- \frac{1}{\left(w \alpha + k^2 - \mu^2_2 + \imath  \Gamma_2\right)}\right],
\\ \nonumber
T_q^{(S)}(w \alpha, k^2)
&=& \frac{ R_S \left(k^2\right)}{(\mu^2_3 - \mu^2_4) - \imath (\Gamma_3 - \Gamma_4)}
\left[\frac{1}{\left(w \alpha + k^2 - \mu^2_3 + \imath  \Gamma_3\right)}
- \frac{1}{\left(w \alpha + k^2 - \mu^2_4 + \imath  \Gamma_4\right)}\right]. 
\end{eqnarray}
It seems that
specifying  $R_{U,S}$ cannot be done unambiguously. We
postpone investigating this problem to the future while in the present paper we
use $R_{U,S}$ defined as follows:

\begin{equation}\label{rus}
R_U = \lambda_U \left(\frac{k^2}{k^2 + \mu^2_U}\right)^{1 + \eta},~~~
R_S = \lambda_S \left(\frac{k^2}{k^2 + \mu^2_S}\right)^{1 + \eta},
\end{equation}
where  $\lambda_{U,S}$ and $\mu^2_{U,S},~(\mu^2_{U,S} > 0)$ are independent parameters, though we think that 
$R_U$ and $R_S$ could coincide which would diminish the number of free parameters.  
It is easy to check now that the expressions for $T_q^{(U)}, T_q^{(S)}$ introduced in Eq.~(\ref{tussud}) obey
the condition of IR stability in Eq.~(\ref{tqir}) with arbitrary $\eta$. In contrast, the value of the UV parameter $\chi$
(introduced in Eqs.~(\ref{tqunpol},\ref{tqspinl}) to
guarantee UV stability)  
is now fixed: $\chi = 1$ in Eq.~(\ref{tussud}). 
%So, our model needs at least two
%resonances. One-resonance model corresponds to $\chi = 0$ which violates the UV stability of the amplitudes $A_q$. 
Eqs.~(\ref{tus},\ref{tussud}) are suggested for invariant amplitudes $T_q^{(U,S)}$ in Basic factorization. Reducing
Basic factorization to $k_T$ -factorization converts $T_q^{(U,S)}$ into new amplitudes $\widetilde{T}_q^{(U,S)}$. They are
obtained from $T_q^{(U,S)}$ by integrating them with respect to $\alpha$:

\begin{equation}\label{tusktgen}
\widetilde{T}_q^{(r)} (\beta, k^2_{\perp}) = \int_0^{\alpha_{\max}} d \alpha T^{(r)}(\alpha,k^2 ),
\end{equation}
where $r = U,S$. The upper limit of integration, $\alpha_{\max}$ should obey Eq.~(\ref{abk}), so we choose

\begin{equation}\label{amax}
\alpha_{\max} \approx k^2_{\perp}/(w \beta).
\end{equation}
According to Eq.~(\ref{abk}),
%we can neglect $w \alpha \beta$ compared to $k^2_{\perp}$, so that
$k^2 \approx -k^2_{\perp}$. The integration leads to
the following expression for $\widetilde{T}_q^{(j)} $ (see Appendix D for detail):
\begin{eqnarray}\label{tuskt}
\widetilde{T}_q^{(U)} (\beta, k^2_{\perp}) &\approx& \frac{1}{2} R_U \left(k^2_{\perp}\right)
\left[\frac{1}{k^2_{\perp}(1 - \beta)/\beta -\mu^2_1 + \imath  \Gamma_1 }
+ \frac{1}{k^2_{\perp}(1 - \beta)/\beta -\mu^2_2 + \imath  \Gamma_2 }\right]
+ \Delta \widetilde{T}_q^{(U)},
\\ \nonumber
\widetilde{T}_q^{(S)} (\beta, k^2_{\perp}) &\approx& \frac{1}{2} R_S \left(k^2_{\perp}\right)
\left[\frac{1}{k^2_{\perp}(1 - \beta)/\beta -\mu^2_3 + \imath  \Gamma_3 }
+ \frac{1}{k^2_{\perp}(1 - \beta)/\beta -\mu^2_4 + \imath  \Gamma_4 }\right]
+ \Delta \widetilde{T}_q^{(S)},
\end{eqnarray}
where $\Delta \widetilde{T}_q^{(U)}$ and $\Delta \widetilde{T}_q^{(S)}$ are

\begin{eqnarray}\label{deltat}
\Delta \widetilde{T}_q^{(U)} = \frac{1}{(\mu^2_1 - \mu^2_2) + \imath (\Gamma_1 - \Gamma_2)}
\ln \left(\frac{k^2_{\perp} + \mu^2_1 - \imath \Gamma_1}{k^2_{\perp}+ \mu^2_2 - \imath \Gamma_2}\right),
\\ \nonumber
\Delta \widetilde{T}_q^{(S)} = \frac{1}{(\mu^2_3 - \mu^2_4) + \imath (\Gamma_3 - \Gamma_4)}
\ln \left(\frac{k^2_{\perp} + \mu^2_3 - \imath \Gamma_3}{k^2_{\perp}+ \mu^2_4 - \imath \Gamma_4}\right).
\end{eqnarray}

They depend on $k_{\perp}$ very slowly and they can be neglected at large $k^2_{\perp}$.

\subsection{Primary quark distributions}

The Optical theorem relates the $s$-channel imaginary parts of $T^{(U,S)}$ and $\widetilde{T}^{(U,S)}$ to
the primary quark distributions $\Psi_{U,S}$ in Basic factorization and to unintegrated (or ) quark distributions $\Phi_{U,S}$
in $k_T$ -factorization respectively.
So applying the Optical theorem, we obtain the following expression for the primary quark distribution $\Psi_r$
in Basic factorization:
\begin{eqnarray}\label{psius}
\Psi_U (w \alpha, k^2) &=& \frac{1}{\pi} \frac{R_U\left(k^2\right)}{(\mu^2_1 - \mu^2_2)}
\left[\frac{\Gamma_1}{\left(w \alpha + k^2 - \mu^2_1\right)^2 + \Gamma^2_1}
- \frac{\Gamma_2}{\left(w \alpha + k^2 - \mu^2_2\right)^2 + \Gamma^2_2}\right],
\\ \nonumber
\Psi_S (w \alpha, k^2) &=& \frac{1}{\pi} \frac{R_S\left(k^2\right)}{(\mu^2_3 - \mu^2_4)}
\left[\frac{\Gamma_3}{\left(w \alpha + k^2 - \mu^2_3\right)^2 + \Gamma^2_3}
- \frac{\Gamma_4}{\left(w \alpha + k^2 - \mu^2_4\right)^2 + \Gamma^2_4}\right].
\end{eqnarray}

and a similar expression for the primary quark distribution $\Phi_{U,S}$ in $k_T$ -factorization:

\begin{eqnarray}\label{phius}
\Phi_U (\beta, k^2_{\perp}) &=& \frac{1}{\pi}  R_U\left(k^2_{\perp}\right)
\left[\frac{\Gamma_1}{\left(k^2_{\perp}(1-\beta)/\beta - \mu^2_1\right)^2 + \Gamma^2_1}
+ \frac{\Gamma_2}{\left(k^2_{\perp}(1-\beta)/\beta - \mu^2_2\right)^2 + \Gamma^2_2}\right],
\\ \nonumber
\Phi_S (\beta, k^2_{\perp}) &=& \frac{1}{\pi}  R_S\left(k^2_{\perp}\right)
\left[\frac{\Gamma_3}{\left(k^2_{\perp}(1-\beta)/\beta - \mu^2_3\right)^2 + \Gamma^2_3}
+ \frac{\Gamma_4}{\left(k^2_{\perp}(1-\beta)/\beta - \mu^2_4\right)^2 + \Gamma^2_4}\right].
\end{eqnarray}

Obviously, the expressions in Eqs.~(\ref{psius},\ref{phius}) are of the Breit-Wigner type.
%Now let us proceed to the primary integrated parton distributions $\phi_{U,S}$ which are used
%in Collinear factorization.
Substituting Eq.~(\ref{phius}) in Eq.~(\ref{dkt}) and integrating over $k^2_{\perp}$,
we arrive at the quark parton distribution $D_j^{(col)}$ in $k_T$ -factorization,
where the non-perturbative contributions i.e. the unintegrated parton
distributions are specified:

\begin{eqnarray}\label{duskt}
D_U^{(k_T)} (x, q^2_{\perp}) = \frac{1}{\pi}\int_x^1
\frac{d \beta}{\beta}
\int_0^r
\frac{d k^2_{\perp}}{k^2_{\perp}}
D_U^{(pert)} (x/\beta, k^2_{\perp}/q^2_{\perp})
R_U\left(k^2_{\perp}\right)
\\ \nonumber
\left[\frac{\Gamma_1}{\left(k^2_{\perp}(1-\beta)/\beta - \mu^2_1\right)^2 + \Gamma^2_1}
+ \frac{\Gamma_2}{\left(k^2_{\perp}(1-\beta)/\beta - \mu^2_2\right)^2 + \Gamma^2_2}\right],
\\ \nonumber
D_S^{(k_T)} (x,q^2_{\perp}) = \frac{1}{\pi}\int_x^1
\frac{d \beta}{\beta}
\int_0^r
\frac{d k^2_{\perp}}{k^2_{\perp}}
D_S^{(pert)} (x/\beta, k^2_{\perp}/q^2_{\perp})
\\ \nonumber
R_S \left(k^2_{\perp}\right)
\left[\frac{\Gamma_3}{\left(k^2_{\perp}(1-\beta)/\beta - \mu^2_3\right)^2 + \Gamma^2_3}
+ \frac{\Gamma_4}{\left(k^2_{\perp}(1-\beta)/\beta - \mu^2_4\right)^2 + \Gamma^2_4}\right],
\end{eqnarray}
where $r$ is defined in Eq.~(\ref{dkt}).
Let us consider the $k_{\perp}$ -dependence in Eqs.~(\ref{phius},\ref{dkt})
%for the parton distributions in $k_T$ -factorization
in more detail.
Obviously, the structures of expressions for $D_U^{(k_T)}$ and $D_S^{(k_T)}$ (or $\Phi_u$ and $\Phi_S$)
are quite similar, so
we consider $D_U^{(k_T)}$ only. Then, the expression in the squared brackets in Eq.~(\ref{dkt}),
i.e. $\Phi_U$ of Eq.~(\ref{phius}),
is symmetric with respect to replacement $1 \rightleftharpoons 2$.  Each term in the
parentheses  has a
peaked form, with maximums at $k^2_{\perp} = \mu^2_{1,2}$. The less $\Gamma_{1,2}$,
the sharper the peaks are. We remind that $R_{U,S} \sim (k^2_{\perp})^{1+\eta}$
at small $k^2_{\perp}$. By definition, see Eq.~(\ref{mu}), $\mu^2_{1,2} = M^2_{1,2} - p^2$,
so they can be either positive or negative while $k^2_{\perp}$ cannot be negative. In
any case the both terms in $\Phi_U$ and $\Phi_S$ contribute to $D_{U,S}^{(k_T)}$ but a result of
interference of the two peaks depends on values of the parameters.
There are possible three particular cases:

\textbf{Case} \textbf{(i):} both $\mu^2_1$ and $\mu^2_2$ are positive.

In this the both maximums are within the integration region of Eq.~(\ref{dkt}) and interference of the two
peaks generates various forms of $\Phi_U (\beta, k^2_{\perp})$ ranging from the
picture with two isolated peaks to
 a kind of plateau, depending on values of $\Gamma_{1,2}$.

\textbf{Case} \textbf{(ii):} $\mu^2_1 >0$ and $\mu^2_2 < 0$ or vice versa.

Here the peak from the first term in Eq.~(\ref{phius}) combines with a tail of the
contribution of the second term whose maximum is beyond the integration region of
Eq.~(\ref{dkt}). The resulting picture has a resemblance to the dual
model combining a resonant and a constant term.

\textbf{Case} \textbf{(iii):} both $\mu^2_1$ and $\mu^2_2$ are negative.

The both maximums now are out of the integration region, so tails of the peaks,
taken by themselves, generate a form slow decreasing with growth of $k^2_{\perp}$.
However, this slope is affected by an impact of $R_U$. We remind that $R_U = 0$ at $k^2_{\perp} = 0$.

\subsection{Primary quark distributions in Collinear factorization}

Performing integration over $k^2_{\perp}$ in Eq.~(\ref{dkt}), we arrive at the parton distributions
$D_j^{(col)}$ in Collinear factorization. Presuming that parameter $\Gamma_j$ is small, we write the
result of the integration in the following form (see Appendix C for detail):

\begin{eqnarray}\label{ducol}
D_U^{(col)}  (x, q^2_{\perp}) \approx  \int_x^1 \frac{d \beta}{\beta} D_U^{(pert)} (x/\beta, q^2_{\perp}/\mu^2_1) \phi_U (\beta, \mu^2_1)
+  \int_x^1 \frac{d \beta}{\beta} D_U^{(pert)} (x/\beta, q^2_{\perp}/\mu^2_2) \phi_U (\beta,\mu^2_2),
\end{eqnarray}
with

\begin{eqnarray}\label{phicol}
\phi_U (\beta, \mu^2_1) &\approx& \frac{1}{\pi}\int_{\Omega_1} \frac{d k^2_{\perp}}{k^2_{\perp}}
\frac{R_U\left(k^2_{\perp}\right) \Gamma_1}{\left(k^2_{\perp}(1-\beta)/\beta - \mu^2_1\right)^2 + \Gamma^2_1},
\\ \nonumber
\phi_U (\beta, \mu^2_2) &\approx& \frac{1}{\pi} \int_{\Omega_2} \frac{d k^2_{\perp}}{k^2_{\perp}}
\frac{R_U\left(k^2_{\perp}\right) \Gamma_2}{\left(k^2_{\perp}(1-\beta)/\beta - \mu^2_2\right)^2 + \Gamma^2_2},
\end{eqnarray}
where the integration regions $\Omega_1 = \Omega'_1 \bigcap [0, w]$ and $\Omega_1 = \Omega'_2 \bigcap [0, w]$,
with the subregions $\Omega'_1, \Omega'_2$ being located around the maximums of the peaks.  Formally,
the both terms in Eq.~(\ref{phicol}) contribute to $\phi_U$ at any signs of $\mu^2_1, \mu^2_2$,
but in the limit of sharp peaks these contributions have different weights:  At $\mu^2_1>0, \mu^2_2 > 0$
the both terms contribute equally:

\begin{equation}\label{phi1}
\phi_U \approx R_U \left(\mu^2_1 \beta/(1- \beta)\right)/\mu^2_1 +
R_U \left(\mu^2_2 \beta/(1- \beta)\right)/\mu^2_2 + O(\Gamma_1, \Gamma_2).
\end{equation}
Mostly
the first term contributes, when $\mu^2_1>0, \mu^2_2 < 0$:

\begin{equation}\label{phi2}
\phi_U \approx R_U \left(\mu^2_1 \beta/(1- \beta)\right)/\mu^2_1 + O(\Gamma_1).
\end{equation}
 and vice versa. Finally, at  $\mu^2_1, \mu^2_2 < 0$ only tails of the both peaks contribute and therefore $\phi_U$
 is small and flat compared to the previous cases:

 \begin{equation}\label{phi3}
 \phi_U \approx const.
 \end{equation}

(see Appendix E for detail).
 When $\mu^2_1>0, \mu^2_2 > 0$

\begin{equation}\label{phiconv}
\varphi_U (\omega,\mu^2) = \int_x^1 \frac{d \beta}{\beta} \beta^{\omega}
\left[
%e^{h_U(\omega)\ln(\mu^2/\mu^2_1)}
%\frac{R_U \left(\mu^2_1 \beta/(1- \beta)\right)}{\mu^2_1}
E(\omega, \mu^2,\mu^2_1)
R_U \left(\mu^2_1 \beta/(1- \beta)\right) \mu^{-2}_2
+
%e^{h_U(\omega)\ln(\mu^2/\mu^2_2)}
%\frac{R_U \left(\mu^2_2 \beta/(1- \beta)\right)}{\mu^2_2}
E(\omega, \mu^2,\mu^2_2)
R_U \left(\mu^2_2 \beta/(1- \beta)\right) \mu^{-2}_2
\right],
\end{equation}
\begin{equation}\label{rumu}
R_U \left(\mu^2_j \beta/(1- \beta)\right) \mu^{-2}_j = \lambda_U \beta.
\end{equation}
Combining Eqs.~(\ref{phiconv}) and (\ref{rumu}), integrating over $\beta$ and remembering that
at small $x$ essential values of $\omega$ are small leads to the following expression for $\varphi_U (\omega, \mu^2)$
(see Appendix E for detail):

\begin{eqnarray}\label{phimu}
\varphi_U (\omega,\mu^2) &=& \int_x^1 \frac{d \beta}{\beta} \beta^{\omega +1}
\left[\frac{\lambda_U}{\mu^2_1}
E(\omega,\mu^2,\mu^2_1)
%e^{h_U(\omega)\ln(\mu^2/\mu^2_1)}
+ \frac{\lambda_U}{\mu^2_1}
E(\omega,\mu^2,\mu^2_2)
%e^{h_U(\omega)\ln(\mu^2/\mu^2_2)}
\right]
\\ \nonumber
&\approx&
%\frac{1}{1 + \omega} \left[
\frac{\lambda_U}{\mu^2_1}
E(\omega,\mu^2,\mu^2_1)
%e^{h_U(\omega)\ln(\mu^2/\mu^2_1)}
+ \frac{\lambda_U}{\mu^2_2}
E(\omega,\mu^2,\mu^2_2)
%e^{h_U(\omega)\ln(\mu^2/\mu^2_2)}
%\right]
.
\end{eqnarray}

%The next step is to specify $R_U$.

\section{Conclusion}

In the present paper we have considered the quark-hadron scattering amplitudes and distributions of
polarized and unpolarized quarks
in hadrons in the framework of the factorization concept where the both amplitudes and distributions are
expressed through convolutions of the perturbative and non-perturbative components.
We began with considering the quark-hadron amplitudes in Basic factorization where integration over momenta of
connecting partons runs over the whole phase space and obtained the conditions for
the factorization convolution to be stable both in IR and UV regions. Then we demonstrated how to reduce Basic
factorization to $K_T$- and Collinear factorizations. We suggested a Resonance Model for non-perturbative contributions
to the unpolarized and spin-dependent parton-hadron scattering amplitudes. This model is based on the
simple argumentation: after emitting an active quark by a hadron, the remaining colored quark-gluon state
cannot be stable and therefore it can be described by quasi-resonant expressions. We needed at least two
resonances  in Basic factorization and this remained true when Basic factorization
  was reduced to $K_T$ -factorization.
  Applying the Optical theorem to the Resonance Model provided us first with the expressions of the
Breit-Wigner type for non-perturbative (primary) contributions to the quark distributions
in Basic and $K_T$ -factorizations and then, after one more reduction, to the parton distributions in Collinear factorization.
To conclude, let us notice that the Resonance Model can also be used for analysis of the non-singlet
components of the DIS structure functions.

\acknowledgements {We are grateful to B.L.~Webber for interesting discussions. S.I.~Troyan is supported by Russian Science Foundation, Grant No.
14-22-00281.

\appendix

\section{Amplitude for the forward Compton scattering off a gluon in the box approximation}

\begin{equation}\label{mg}
M_{\mu\nu\lambda\rho} = (t_lt_r) \frac{e^2 \alpha_s}{8 \pi^2} w \int d \beta' d \alpha'
d k'^2_{\perp} \left[ M_{\mu\nu\lambda\rho}^{(1)} +  M_{\mu\nu\lambda\rho}^{(2)} +  M_{\mu\nu\lambda\rho}^{(3)} \right]
\end{equation}
\begin{eqnarray} \label{mg123}
 M_{\mu\nu\lambda\rho}^{(1)} = \frac{Tr \left[\gamma_{\nu} \left(\hat{q} + \hat{k'}\right)\gamma_{\mu}\hat{k'}
\gamma_{\lambda}\left(\hat{k'}- \hat{k}\right)\gamma_{\rho}\hat{k'}\right]}{k'^2 k'^2 (q+k')^2 (k'-k)^2},
\\ \nonumber
M_{\mu\nu\lambda\rho}^{(2)} = \frac{Tr \left[\gamma_{\nu} \left(\hat{q} + \hat{k'}\right)\gamma_{\mu}\hat{k'}
\gamma_{\rho}\left(\hat{k'} + \hat{k}\right)\gamma_{\lambda}\hat{k'}\right]}{k'^2 k'^2 (q+k')^2 (k' + k)^2},
\\ \nonumber
M_{\mu\nu\lambda\rho}^{(3)} = \frac{Tr \left[\gamma_{\nu} \left(\hat{q} + \hat{k'} - \hat{k}\right)
\gamma_{\rho}\left(\hat{k'} - \hat{k}\right)\gamma_{\mu} \hat{k'}\gamma_{\lambda}
\left(\hat{k'} + \hat{q}\right)\right]}
{k'^2 (q+k')^2 (k' - k)^2 (q + k' - k)^2}
 \end{eqnarray}

\section{Projection operators for forward Compton amplitudes}

The conventional of dealing with the forward Compton scattering amplitude $A_{\mu\nu}$ is, in the first place, to
simplify their tensor structure. To this end, $A_{\mu\nu}$ is represented as an expansion
of $A_{\mu\nu}$ into the series of simpler tensors, each
multiplied by an invariant amplitude. Such tensors are called projection operators. Through the Optical
theorem the invariant amplitudes are related to the DIS structure functions.

In the case of the unpolarized Compton scattering such an expansion looks as follows:

\begin{equation}\label{expunpol}
A_{\mu\nu} = P_{\mu\nu}^{(1)} A_1 +  P_{\mu\nu}^{(2)} A_2,
\end{equation}
where

\begin{equation}\label{p12unpol}
P_{\mu\nu}^{(1)} = - g_{\mu\nu} + q_{\mu}q_{\nu}/q^2,~~
P_{\mu\nu}^{(2)} = (1/pq) \left(p_{\mu} - q_{\mu} (pq/q^2) \right)\left(p_{\nu} - q_{\nu} (pq/q^2) \right)
\end{equation}
are the projection operators and $A_1,~A_2$ are invariant amplitudes. According to the Optical theorem

\begin{equation}\label{f12}
F_1 = \frac{1}{\pi} \Im A_1,~F_2 = \frac{1}{\pi} \Im A_2.
\end{equation}

Similarly, for the polarized Compton scattering

\begin{equation}\label{expspin}
A_{\mu\nu} = P_{\mu\nu}^{(3)} A_3 +  P_{\mu\nu}^{(4)} A_4,
\end{equation}
where

\begin{equation}\label{p34spin}
P_{\mu\nu}^{(3)} = \imath \epsilon_{\mu\nu\lambda\rho} M q_{\lambda} S_{\rho},~
P_{\mu\nu}^{(4)} = \imath \epsilon_{\mu\nu\lambda\rho} M q_{\lambda}
\left[ S_{\rho} - p_{\rho} \left(qS/qp\right)\right],
\end{equation}
with $M$ and $S$ being the hadron mass and spin respectively,
and $A_{3,4}$ are spin-dependent invariant amplitudes. The Optical theorem states that

\begin{equation}\label{g12}
g_1 = \frac{1}{\pi} \Im A_3,~ g_2 = \frac{1}{\pi} \Im A_4.
\end{equation}

All operators $P_{\mu\nu}^{(n)}$ respect the electromagnetic current conservation: $q_{\mu}P_{\mu\nu}^{(n)} = q_{\nu}P_{\mu\nu}^{(n)}=0$.
It is convenient to introduce the longitudinal, $S^{||}$ and transverse, $S^{\perp}$ components of the spin, so that
$S^{\perp}p = S^{\perp}q = 0$ and $S^{||}_{\rho} = p_{\rho} (qS/pq)$. In such terms Eq.~(\ref{expspin}) can be written as follows:

\begin{equation}\label{expspin1}
A_{\mu\nu} =  \imath \epsilon_{\mu\nu\lambda\rho} M q_{\lambda} \left[ S^{||}_{\rho} A_3 +
S^{\perp}_{\rho} \left(A_3 + A_4\right)\right] \equiv
\imath \epsilon_{\mu\nu\lambda\rho} M q_{\lambda} \left[ S^{||}_{\rho} A^{||} +
S^{\perp}_{\rho} A^{\perp} \right].
\end{equation}

This expression is useful for practical attributing different terms in the spin-dependent $A_{\mu\nu}$ to
proper invariant amplitudes. In the unpolarized case one can use the simple rule: expressions $\sim g_{\mu\nu}$
contribute to $A_1$ while expressions $\sim p_{\mu}p_{\nu}/pq$ form $A_2$. In contrast, the gauge invariance admits
adding arbitrary terms $\sim q_{\mu},~q_{\nu}$.

\section{Convolutions involving the Breit-Wigner formula}

Let us consider the following convolution:

\begin{equation}\label{bwconv}
F = \frac{1}{\pi}\int_{-\infty}^{\infty} dx f(x) \frac{\Gamma}{(x-x_0)^2 + \Gamma^2}.
\end{equation}
Replacing $x$ by $t$, with $t = (x - x_0)/\Gamma$, we convert Eq.~(\ref{bwconv}) into

\begin{equation}\label{bwconvt}
F = \frac{1}{\pi}\int_{-\infty}^{\infty} dt f(t \Gamma + x_0) \frac{\Gamma}{t^2 + 1}.
\end{equation}
At small $\Gamma$, we can expand $f(t \Gamma + x_0)$ in the power series and retain several terms:

\begin{equation}\label{fser}
f(t \Gamma + x_0) = f(x_0) + f'(x_0) t \Gamma + O(\Gamma^2).
\end{equation}

Substituting Eq.~(\ref{fser}) in (\ref{bwconvt}) and integrating (\ref{bwconvt}) yields

\begin{equation}\label{bwrezt}
F = f(x_0) + O(\Gamma).
\end{equation}

The first term in Eq.~(\ref{bwrezt}) corresponds to the well-known representation of the $\delta$ -function:

\begin{equation}\label{delta}
\frac{1}{\pi} \lim_{\epsilon \to 0}  \frac{\epsilon}{x^2 + \epsilon^2} = \delta(x)
\end{equation}

\section{Integration in Eq.~(\ref{tusktgen})}

A generic expression to integrate can be written as

\begin{equation}\label{tktgen}
\widetilde{T} = \int_0^{\alpha_{\max}} \frac{d \alpha}{(\alpha - A)(\alpha - B)} = \frac{1}{(A-B)}
\int_0^{\alpha_{\max}} d \alpha \left[\frac{1}{\alpha - A} - \frac{1}{\alpha - B}\right]
= \frac{1}{(A-B)} \left[\ln \left(\frac{\alpha_{\max} - A}{\alpha_{\max} - B}\right)
- \ln \left(\frac{A}{B}\right)\right].
\end{equation}
Assuming that $\alpha_{\max} \gg A,B \gg |A-B|$ allows us to
expand the logarithms into the power series and retain the first
terms only:

\begin{equation}\label{tkt}
\widetilde{T} \approx \frac{1}{2} \left[\frac{1}{\alpha_{\max} - A} + \frac{1}{\alpha_{\max} - B}
\right] - \frac{1}{(A-B)}\ln \left(A/B\right).
\end{equation}
We have written Eq.~(\ref{tkt}) in the symmetrical form with respect to $A,B$ because Eq.~(\ref{tktgen}) has this feature.
%The integration yields
%\begin{equation}\label{tktln}
%\widetilde{T} = \frac{1}{(A-B)} \left[\ln \left(\frac{\alpha_{\max} - A}{\alpha_{\max} - B}\right)
%- \ln \left(A/B\right)\right]
%\end{equation}

\section{Evolving the factorization scale in Collinear factorization}

 Using the Mellin transform, we can rewrite Eq.~(\ref{dcol}) as follows:
% with the arbitrary factorization scale $\mu$:

\begin{equation}\label{dmellin}
D^{(col)}(x, q^2_{\perp}) = \int_{- \imath \infty}^{\imath \infty} \frac{d \omega}{2 \pi \imath}
x^{- \omega} C_U(\omega)
E(\omega, q^2, \mu^2) \phi (\omega,\mu^2),
\end{equation}
where the primary quark distribution is fixed at the scale $\mu$, with $\mu^2 < q^2_{\perp}$.
 $E (\omega, q^2_{\perp}, \mu^2)$ is a generic notation for an operator evolving the
 distribution $\phi (\omega,\mu^2)$ from the factorization scale $\mu^2$ to $q^2_{\perp}$,
 while $C (\omega)$ is responsible for the $x$ -evolution.
 Choosing a scale $\widetilde{\mu}$ such that
 $\mu^2 <\widetilde{\mu} < q^2_{\perp}$  and representing $E (\omega, q^2_{\perp}, \mu^2)$
 as
 \begin{equation}\label{mutildemu}
 E (\omega, q^2_{\perp}, \mu^2) = E (\omega, q^2_{\perp}, \widetilde{\mu}^2)E (\omega, \widetilde{\mu}^2, \mu^2),
 \end{equation}
we bring $D^{(col)}$ to the conventional form
\begin{equation}\label{dmellinconv}
D^{(col)}(x, q^2_{\perp}) = \int_{- \imath \infty}^{\imath \infty} \frac{d \omega}{2 \pi \imath}
x^{- \omega} C_U(\omega)
%e^{h_U(\omega)\ln(q^2/\mu^2)} \varphi_U (\omega,\mu^2),
E(\omega, q^2, \widetilde{\mu}^2) \varphi (\omega,\widetilde{\mu}^2),
\end{equation}
where
\begin{equation}\label{psipsitildemel}
\varphi(\omega,\widetilde{\mu}^2) = E (\omega, \widetilde{\mu}^2, \mu^2) \phi (\beta, \mu^2),
\end{equation}
which corresponds to Eq.~(\ref{phimutilde}).
Actually,  $\varphi(\omega,\widetilde{\mu}^2)$ is the conventional parton distribution in the $\omega$ -space (momentum space).
 It is fixed at an arbitrary scale $\widetilde{\mu}^2$  and
 related to the standard integrated distribution $\delta q (x,\widetilde{\mu}^2)$ by the Mellin transform.
 The evolution operator $E(\omega, q^2, \mu^2)$ is expressed in different terms, depending on the
perturbative approach in use. For instance, in LO DGLAP with fixed $\alpha_s$ it is given by

\begin{equation}\label{edglapfix}
E = \exp \left[\alpha_s \gamma_0 (\omega) \ln (q^2/\mu^2) \right],
\end{equation}
with $\gamma_0$ being the LO DGLAP anomalous dimension and $C_F = 4/3$.
When in LO DGLAP $\alpha_s$ is running and the standard parametrization
$\alpha_s = \alpha_s \left(k^2_{\perp}\right)$ is used in the Feynman graphs, Eq.~(\ref{edglapfix}) is changed by
\begin{equation}\label{edglap}
E = \left(\frac{q^2}{\mu^2}\right)^{\gamma_0/b},
\end{equation}
with $b$ being the first coefficient of the $\beta$ -function. The parametrization
$\alpha_s = \alpha_s \left(k^2_{\perp}\right)$ should not be used at small $x$ (see Ref.~\cite{egtsum}). When
it is replied by the appropriate parametrization and when the total resummation of
the leading logarithms is done, Eq.~(\ref{edglap})  is
replaced by

\begin{equation}\label{eegt}
E = \exp \left[h(\omega) \ln(q^2/\mu^2)\right],
\end{equation}
where $h(\omega)$ is a new anomalous dimension. It accounts for the total resummation of the
leading double-logarithmic contributions and running QCD coupling effects (see Ref.~\cite{egt} and overview \cite{egtsum} for detail).

\end{document}